\documentclass{ws-procs975x65}

\begin{document}


\title{GENERAL RELATIVITY, DIFFERENTIAL GEOMETRY, AND UNITARY 
THEORIES IN THE WORK OF MIRA FERNANDES}

\author{JOS\'E P. S. LEMOS}

\address{Centro Multidisciplinar de Astrof\'\i sica - CENTRA,
Departamento de F\'\i sica,\\ Instituto Superior T\'ecnico - IST,
Universidade Universidade T\'ecnica de Lisboa - UTL,\\ Av. Rovisco
Pais 1, 1049-001 Lisboa, Portugal\\ E-mail: joselemos@ist.utl.pt}

\begin{abstract}
An analysis of the work of Mira Fernandes on unitary theories is
presented. 
\end{abstract}


\bodymatter
\section{Introduction}\label{sec:intro}
Aureliano de Mira Fernandes, born in 1884 in Portugal, was professor
of Differential and Integral Calculus and Rational Mechanics, at
Instituto Superior T\'ecnico, from its foundation in 1911, onwards,
until his retirement. He was also professor of Mathematical Analysis
at the Instituto Superior de Ci\^encias Econ\'omicas e Financeiras,
what is now the Instituto Superior de Economia e Gest\~ao.  Mira
Fernandes, by formation, was a mathematician.  His Doctoral
dissertation in 1911, supervised by Sid\'onio Pais, was on ``Galois
theory'', defended when he was 27 years old.  From his dissertation to
1924 there are no publications. From 1924 onwards there are many
publications, more than eighty, on several subjects, namely, group
theory, differential geometry, unitary theories and rational
mechanics. The most important papers were published in Rendiconti
della Accademia dei Lincei, due to his friendship with Levi-Civita,
the great Italian mathematical physicist. After Levi-Civita's
compulsory retirement, he published mainly in Portuguese journals.  He
also corresponded with Elie Cartan. Cartan in his work ``Les espaces
de Finsler'' writes ``It is after an exchange of letters with
M. Aurelio (sic) de Mira-Fernandes, that I have perceived of the
possibility of this simplification''.  He was a member of the Lisbon
Academy of Sciences from 1928 onwards. In 1932 he proposed Levi-Civita
and Einstein to be foreign members of the Academy, a proposal accepted
immediately by the President of the Academy, Egas Moniz, the future
Nobel prize in medicine. These proposals were apt, since these two
figures were pioneers in differential and Riemannian geometry, general
relativity and unitary theories, areas for which Mira Fernandes
devoted a great part of his scientific life (see \cite{lemos} for the
complete version of this article).

\section{The scientific context of unitary theories}
\label{scientificcontext}

The idea of unification in physics is an old one, with Mie's 
ideas being in the forefront in the 1912s.  General relativity
with its beautiful geometric structure, changed the picture from 1916
onwards. It put the gravitational field in a special relativity
framework.  However, it left electromagnetism out.  Now, electricity
and magnetism had been unified into electromagnetism using special
relativity and a spacetime arena.  Thus, one might
argue, gravity (general relativity) and electromagnetism should be
unifiable in a unitary theory using a special world background as
the new arena. This was advocated by many, notably by Eddington.
What this world background could be was left imprecise. This rationale
works if general relativity is a field theory, on the same footing of
electromagnetism. But even this is controversial.

The first attempt to unify gravitation and electromagnetism was
proposed by Weyl$\,$\cite{weyl1918}$\,$. In this theory the
electromagnetic potential is introduced as a geometrical quantity
which determines the transport law of a length scale.  Weyl was able
to reproduce Einstein's and Maxwell's equation within a single
scheme. However, when confronted with observations the theory does not
hold, atoms arriving at the earth from the cosmos would have different
physical properties, as pointed out by Einstein.  In spite of this
demolishing problem, Weyl's idea of gauging was one of the most
fruitful ideas in the history of physics in the context of quantum
mechanics and electromagnetism.  In any case the door to unification
schemes was open. There is a Brazilian saying that says ``Where one ox
passes a herd of oxen passes''. It applies neatly here.  The next ox
to pass was Eddington's theory$\,$\cite{eddingtonpaper}$\,$. Eddington
set forward the idea that, perhaps, the connection
$\Gamma^\lambda_{\mu\nu}$ is the primary quantity, rather than the
metric $g_{\mu\nu}$ 
itself and partially developed a theory. Einstein in between
the years of 1923 and 1925 fiddled with the theory but could not
progress.  

Further ideas on connections and parallel transport from
mathematicians were also used to propose physical theories of
gravitation and electromagnetism.
A general connection $\Gamma$ (dropping the indices) has a metric
part, a homothetic part as in Weyl's theory, and a torsion, which is
the antisymmetric part of the connection.  Thus, besides the
Riemann-Christoffel curvature, one gets a homothetic curvature, and a
torsion curvature. Unitary theories, that tried to unify the
gravitational and electromagnetic fields used one or all these
connections and curvatures.  To see some of these, 
see$\,$\cite{tonnelatbook1965goennerreview} for reviews and precise
citations, see also$\,$\cite{lemos}$\,$.
Another idea on connections that sprang at the time, and is seldom
mentioned, is that the manifold can see a connection $\Gamma$ for
contravariant vectors $v$ and a different connection
$\Gamma\,'$ for covariant vectors $u$. Thus, for each
connection, $\Gamma$ and $\Gamma\,'$, one gets the usual
Riemann-Christoffel curvature, a torsion curvature, and a homothetic
curvature. These two connections give rise to a new three-index tensor
field $C$ which in turn makes the bridge between the connections
themselves. The field $C$ is defined as the covariant derivative of
the identity tensor $I$. In most theories, this $C$ field was put to
zero, probably because of its apparent lack of physical meaning. As we
will see, not for Mira Fernandes. In the years between 1926 and 1933
he explored some of the proposed theories by adding to them the $C$
field, while trying to physically interpret it.

\section{\small\bf The works of Mira Fernandes on unitary theories of
gravitation and electromagnetism} \label{mirafernandes}

The book ``Foundations of differential geometry of the linear spaces''
(in Portuguese), 1926 is based on Schouten's book of 1924
``Ricci-Kalk\"ul'' (in German) and displays the theory of connections
at the time.  A paper of 1931 in Rendiconti shows some seven new
properties of connections.

In the paper of 1932$\,$\cite{mirapaper1932} Mira Fernandes ventures
into unitary theories. He analyzes Straneo's papers, an Italian
mathematical physicist from the group of Levi-Civita. Straneo
published papers on unitary theories, see his
review$\,$\cite{strnuoci}$\,$.  Straneo's connection that most
interests Mira Fernandes is
$
\Gamma^\alpha_{\beta\gamma}=\{^{\,\,\alpha\,}_{\beta\gamma}\}
+\left(\delta^\alpha_\mu\psi_\nu-\delta^\alpha_\nu\psi_\mu
\right)\,,
\label{straneoconnection}
$
where $\psi_\nu$ is an additional vector field of the theory, to be
equated physically to the electromagnetic potential. This equation is
based on Weyl's connection and generalizes it. However, it does not
have the same mathematical and physical substrata.  Mira Fernandes
supposes that his own connection is invariant by incidence, i.e.,
${C_{\alpha\beta}}^\gamma=C_\alpha\,I_\beta^\gamma$, 
covariant symmetric, and contravariant metric.  The aim of his Note is
to formulate considerations about Straneo's connection, involving the
field $C_\alpha$.  For a four-dimensional spacetime, $n=4$, the
connection, in Mira Fernandes' words, satisfies ``all the conditions
attributed by Straneo to the structure of the physical space''. The
vector $\psi_\alpha$ representing the electromagnetic potential is
such that
$
\psi_\alpha=-\,\frac12 \,C_\alpha
\,,
\label{psiasc}
$
and the torsion $S$ and nonmetricity $Q$ are given by
$
S_\alpha=\frac12\,Q_\alpha=C_\alpha\label{psiasc2}
$.
The field $C$ provides much of the physical and geometrical content of
the theory. 
It is also noted that the connection ``is distinct from
Weyl's since $C\neq0$''.  When
$C_\alpha=C_{,\alpha}$, i.e,  is a gradient, then one
recovers Weyl.
There are other related publications, see$\,$\cite{lemos}$\,$.

\section{Conclusions}
\label{conclusions}

Theories which change Riemannian geometry, as those used by Mira
Fernandes, are not in fashion as theories of unification,
they were reverted to theories of gravitation and spin, the
Einstein-Cartan theories.  Theories that use extra dimensions
to obtain fields in four dimensions, the Kaluza-Klein theories,
were not touched by Mira Fernandes. These ideas are still in use
in supergravity and string theory. The name of such
theories has been changing, unitary theories at first, then unified
field theories, and nowadays theories of everything. Will their fate
be the same as Mie's theory?


\vskip 0.1cm
\noindent
Acknowledgments. This work was partially supported by
FCT - Portugal (CERN/FP/109276/2009 and PTDC/FIS/098962/2008).

\bibliographystyle{ws-procs975x65}
\bibliography{ws-pro-sample}
\end{document}